\documentclass[letter,12pt]{article}
\pdfoutput=1 

\usepackage{jheppub} 

\usepackage[utf8]{inputenc}

\usepackage[russian,english]{babel}
\usepackage[T1,T2A]{fontenc} 

\usepackage{amsmath}
\DeclareMathOperator{\Tr}{Tr}
\usepackage{ upgreek }

\def\bea{\begin{eqnarray}}
\def\eea{\end{eqnarray}}

\def\n{{\rm n}}

\def\z{{e^*}}

\title{Generalized Gibbs Ensemble of 2d CFTs at large central charge in the thermodynamic limit}

\author[a,b]{Anatoly Dymarsky} 
\author[b, c]{and Kirill Pavlenko}

\affiliation[a]{University of Kentucky,\\Lexington, KY, USA 40506\\}
\affiliation[b]{Skolkovo Institute of Science and Technology,\\Skolkovo Innovation Center, Moscow, Russia\\}
\affiliation[c]{Moscow Institute of Physics and Technology, Dolgoprudny 141700, Russia}
\emailAdd{a.dymarsky@uky.edu}
\emailAdd{kirill.pavlenko@skoltech.ru}

\abstract{We discuss partition function of 2d CFTs decorated by higher qKdV charges in the thermodynamic limit when the size of the spatial circle goes to infinity. In this limit the saddle point approximation is exact and at infinite central charge  generalized partition function can be calculated explicitly. We show that leading $1/c$ corrections to free energy can be reformulated as a sum over Young tableaux which we calculate for the first two qKdV charges. Next, we compare generalized ensemble with the ``eigenstate ensemble'' that consists of a single primary state. At infinite central charge the ensembles match at the level of expectation values of local operators for any values of qKdV fugacities. When the central charge is large but finite, for any values of the fugacities the aforementioned ensembles are distinguishable. }

\begin{document} 
\maketitle
\flushbottom


\section{Introduction}
\label{sec:intro}
Conformal Field Theories in two dimensions describe one dimensional systems at criticality, most notably  string theoretic worldsheet. Thermalization of two-dimensional theories came to the focus of attention recently in the context of entanglement spreading following a quantum quench \cite{calabrese2005evolution,calabrese2006time,calabrese2007quantum}. Additional motivation to study 2d theories comes from holographic duality, which relates thermal properties of boundary CFTs to black hole physics in $AdS_3$ \cite{maldacena2003eternal,hartman2013time,fitzpatrick2014universality,fitzpatrick2015virasoro,fitzpatrick2016information,fitzpatrick2016conformal,anous2016black,chen2017numerical,fitzpatrick2017exact,fitzpatrick2017late,faulkner2018probing}.  
Thermal partition function of 2d CFTs on a torus is a well-studied object which exhibits elegant mathematical properties making a connection with the theory of modular forms \cite{green1987superstring,francesco2012conformal}. In the thermodynamic limit, when the size of  spatial circle $\ell$ goes to infinity, torus degenerates into a cylinder and thermal partition function becomes particularly simple. At the level of free energy it only depends on central charge $c$ and inverse temperature $\beta$,
\bea
F={c\,\pi^2 \ell\over 6\, \beta}.
\eea
The peculiar property of conformal field theories in two dimensions is that, in addition to energy, there is an infinite number of local conserved charges associated with quantum KdV algebra  \cite{bazhanov1996integrable,bazhanov1997integrable,bazhanov1999integrable}. Whenever a system has  additional conserved quantity, e.g.~total number of particles, its thermal properties are described by the grand canonical ensemble, with the effective number of particles controlled by the value of chemical potential \cite{mingo2018quantum}. In principle similar logic applies when there are many conserved quantities.  When the system is integrable the number of conserved quantities becomes infinite, raising the question if such a system would even approach thermal equilibrium, and what that equilibrium state might be. These questions have drawn significant attention recently because of experimental realizations of such a scenario \cite{kinoshita2006quantum}. The prevailing view is that integrable system would evolve toward a state described by the Generalized Gibbs Ensemble (GGE) \cite{rigol2007relaxation, calabrese2011quantum,caux2012constructing}, which includes all conserved mutually-commuting charges $Q_k$,
\bea
\label{GGEdef}
Z={\rm Tr}\left(e^{-\sum_k \mu_k Q_k}\right),\qquad Q_1=H,\qquad \mu_1=\beta.
\eea
The value of corresponding chemical potentials, also called fugacities, $\mu_k$ should be chosen such that the ensemble \eqref{GGEdef} would have the same value of conserved charges as the initial state. Applying this logic to the two-dimensional CFTs it is natural to expect that e.g.~following a quantum quench these theories would thermalize to a state which has the same local properties as the GGE decorated by higher qKdV charges \cite{fioretto2010quantum,mandal2015thermalization,sotiriadis2016memory,cardy2016quantum,de2016remarks,bernard2016conformal,alves2017momentum}. This expectation motivates our study of GGE  of two-dimensional theories in this paper. 

We start with a CFT on a spatial circle of length $\ell$, such that Euclidean space-time is parametrized by a holomorphic coordinate $\omega=u+i t$. The latter is related to the conventional coordinate on the plane via $z=e^{{2\pi i\over \ell}\omega}$. In addition to the Hamiltonian 
\bea
\label{H}
H=Q_1=\int_0^\ell du\, T(u),
\eea
conformal theory always  has an infinite tower of mutually commuting qKdV charges \cite{bazhanov1996integrable}
\bea
\label{Q3Q5}
Q_3=\int_0^{\ell} du\, : T(u)^2 :,\qquad Q_5=\int_0^{\ell} du\, : T(u)^3 : + \frac{c+2}{12} : T'(u)^2	:,\qquad  \dots 
\eea
By convention qKdV charges are parametrized by odd numbers $Q_{2k-1}$, $k\geq 1$. We are interested in calculating 
\bea
\label{GGE}
Z={\rm Tr}\left(e^{-\beta H-\sum_{k=2}^\infty \mu_{2k-1}Q_{2k-1}}\right).
\eea
When all $\mu_{2k-1}=0$ for $k\geq 2$,  \eqref{GGE} is the regular partition function on a torus with the modular parameter $\tau=i{\beta\over 2 \pi \ell}$. In particular partition function is invariant under modular transformations of $\tau$. Turning on $\mu_{2k-1}$ would break modular invariance, although  $Z$ still exhibits some interesting properties under modular transformations \cite{maloney1}.  
In this work we focus on thermodynamic limit $\ell\rightarrow \infty$, while $\beta,\mu_{2k-1}$ are kept fixed. Our goal is to calculate the extensive part of free energy $F=\log Z$,
\bea
F={c\,\pi^2 \ell\over 6\, \beta}f(\beta,\mu_{2k-1},c) +o(\ell).
\eea
We show that at infinite central charge $f$ can be expressed in terms of an algebraic relation, while leading $1/c$ corrections can be reformulated as a straightforward sum over Young tableaux which leads to a concise and explicit answer \eqref{otvet}. 

In the second part of the paper we compare local properties of the GGE \eqref{GGE} with those of an individual  energy eigenstate. We find that at infinite central charge local properties of a highly excited Virasoro primary eigenstates exactly match those of the GGE for any values of the fugacities $\mu_{2k-1}$. In this sense heavy primary states can be regarded as thermal at leading order in $c$. At the same time when  central charge is  finite, GGE and a primary eigenstate are always distinguishable in terms of expectation values of local operators. In other words, it is impossible to tune the fugacities $\mu_{2k-1}$ such that the GGE would have the same values of qKdV charges $Q_{2k-1}$ as a heavy CFT primary.  This conclusion is intriguing because it clearly shows that local properties of 2d CFTs upon equilibration are not always governed by the generalized ensemble of the form \eqref{GGE}. Our finding emphasizes that integrable structure of 2d theories may preclude them to thermalize in the sense that local properties of the equilibrated state may lack universality. 


\section{Generalized Gibbs Ensemble in the thermodynamic limit}
The crucial simplification of thermodynamic limit $\ell\rightarrow \infty$ is that saddle point approximation becomes exact. We first illustrate that in the case of the conventional Gibbs ensemble
\bea
\label{Gibbs}
Z(\beta)={\rm Tr}(e^{-\beta H}),\quad H={L_0 - c/24\over \ell}.
\eea
Here $L_n$ is the conventional Virasoro algebra generator related to the stress tensor on the plane,
\bea
T(z)=\sum_{n} {L_{n}\over z^{n+2}},
\eea
while the Hamiltonian of CFT on a circle is given by \eqref{H}.\footnote{The stress tensor on the cylinder $T(u)$ is related to the stress tensor on the plane $T(z)$ via the standard conformal transformation.}
The sum in \eqref{Gibbs} goes over all states in the Hilbert space, 
\bea\label{basis}
|m_i,\Delta\rangle=L_{-m_1}\dots L_{-m_k}|\Delta\rangle,\qquad m_i\geq m_j \ \ {\rm for} \ \ i\geq j, 
\eea
parametrized by the dimension of Virasoro primary $\Delta$ and sets $\{m_i\}$, $\sum_{i=1}^k m_i=n$, arranged in the dominance order.  
Using the degeneracy of $H$, $L_0|m_i,\Delta\rangle=(\Delta+n)|m_i,\Delta\rangle$ the partition function can be represented as a double sum,
\bea
\label{doublesum}
Z(\beta)=\sum_{\Delta} \sum_{n=0} P(n)\, e^{-{\beta\over \ell}(\Delta +n)}.
\eea
In what follows we are only interested in the extensive part of free energy, and therefore we dropped the explicitly $c$-dependent term in $H$. The sum $\sum_{\Delta}$ goes over  all Virasoro primaries including possible multiplicities, and $P(n)$ is the number of integer partitions -- Young tableaux consisting of $n$ elements. For large $n$ \cite{erdos1942elementary},
\bea
P(n)\approx e^{\pi \sqrt{2n\over 3}}
\eea
and the sum over $n$ can be substituted by an integral. Similarly, the sum over $\Delta$ in \eqref{doublesum} can be substituted by an integral multiplied by the density of primaries given by Cardy formula \cite{cardy1986operator}. Combining all together gives 
\bea
\label{Cardy}
Z(\beta)= \int d\Delta\, e^{\pi \sqrt{\frac{2}{3}(c-1)\Delta}} \int dn\, e^{\pi \sqrt{2n\over 3}} e^{-{\beta\over \ell}(\Delta +n)} =e^{F},\quad F={c\, \pi^2\,\ell\over 6\, \beta}. 
\eea
It is easy to see that  in the limit $\ell\rightarrow \infty$ the saddle point approximation is exact, 
\bea
&&{\mathcal L}=\pi \sqrt{\frac{2}{3}(c-1)\Delta}+\pi \sqrt{2n\over 3}-{\beta\over \ell}(\Delta +n),\\
&&\Delta^*= (c-1) \pi^2 \ell^2/ 6\beta^2,\quad n^*=\pi^2 \ell^2/ 6\beta^2 ,\qquad \left.{\partial {\mathcal L}\over \partial \Delta}\right|_{\Delta^*,n^*}=\left.{\partial {\mathcal L}\over \partial n}\right|_{\Delta^*,n^*}=0, \label{saddle}\\
&&F={\mathcal L}(\Delta^*, n^*) = {c\, \pi^2\,\ell\over 6 \beta}.
\eea

\subsection{GGE at Infinite Central Charge}
Our next step is to decorate the partition function by higher qKdV charges,
\bea
Z(\beta,\mu_{2i-1})={\rm Tr}(e^{-\beta H-\sum_k \mu_{2k-1} Q_{2k-1}}).
\eea
We start our analysis with $Q_3$ \eqref{Q3Q5}. The expression for $Q_3$ in terms of Virasoro generators can be found in  \cite{bazhanov1996integrable} (in our case expression for $Q_3$ is different by an overall factor $1/\ell^3$). Using  the explicit form of $Q_3$  we split  it into two parts as follows 
\bea
\label{Q3}
Q_3&=&\hat{Q}_3+\tilde{Q}_3\\  \ell^3 \hat{Q}_3&=&\left(L_0^2 - \frac{c+2}{12} L_0  + \frac{c(5c+22)}{2880} \right),\quad  \ell^3 \tilde{Q}_3=2\sum_{n=1}^\infty L_{-n} L_n.
\eea
The term $\tilde{Q}_3$ is defined such that it annihilates primary states, $\tilde{Q}_3|\Delta\rangle=0$, while the rest is degenerate and depends only on the combination $\Delta+n$. Using scaling with $\ell$ of the saddle point values $\Delta^*\sim \ell^2$ and $n^*\sim \ell^2$ we immediately find that $L_0$-independent term from $\hat{Q}_3$ would give $1/\ell^3$ contribution to free energy, while $- \frac{c+2}{12} L_0$  will contribute as $\sim 1/\ell$. Hence in the thermodynamic limit these terms can be neglected.  Assuming for simplicity that only $\mu_3$ is turned on we get, 
\bea
Z(\beta, \mu_3)&=&\sum_{\Delta} \sum_{n=0} P(n)\, e^{-{\beta \over \ell} (\Delta +n)-{\mu_3\over \ell^3}(\Delta +n)^2+\tilde{\mathcal L}(c,\mu_3/\ell^3,\Delta, n)},\\
e^{\tilde{\mathcal L}(c,\mu_3/\ell^3,\Delta, n)}&\equiv& {1\over P(n)}\sum_{\{m\}=n} \langle m_i,\Delta|e^{-{\mu_3}\tilde{Q}_3}|m_i,\Delta\rangle. \label{summ}
\eea
In \eqref{summ} the sum is over sets $\{m_i\}$ satisfying $\sum_i m_i=n$, i.e.~over the partitions of $n$.

A crucial simplification, which will be justified in the next section, is that $\tilde{Q}_3$ does not contribute to free energy at leading order in  $1/c$ expansion. Hence at infinite central charge one can simply take $\tilde{\mathcal L}$ to be zero,
\bea
Z(\beta, \mu_3)=\int d\Delta \int dn\, e^{\mathcal L},\\ {\mathcal L}=\pi \sqrt{\frac{2}{3}(c-1)\Delta}+\pi \sqrt{2n\over 3}-{\beta\over \ell}(\Delta +n)-{\mu_3\over \ell^3}(\Delta +n)^2. \label{effL}
\eea
Adding higher charges can be done in a similar way. Thus $Q_5$, also given in \cite{bazhanov1996integrable}, can be split into 
\bea
Q_5&=&\hat{Q}_5+\tilde{Q}_5\\
\ell^5 \hat{Q}_5&=&\left(L_0^3- \frac{c+4}{8} L_0^2 + \frac{(c+2)(3c + 20)}{576}L_0  - \frac{c(3c + 14)(7c + 68)}{290304}\right), \\
\ell^5 \tilde{Q}_5 &=& \sum_{\substack{n_1 + n_2 + n_3 =0 \\ n_1 + n_2 + n_3 \neq 0}} :L_{n_1} L_{n_2} L_{n_3}: + \sum_{n = 1}^{\infty} \left( \frac{c+11}{6}n^2 - 1 -\frac{c}{4}  \right)L_{-n} L_n  + \frac{3}{2} \sum_{r=1}^{+\infty} L_{1-2r} L_{2r-1}.  \nonumber
\eea
Normal ordering in $:L_{n_1} L_{n_2} L_{n_3}: $ means that the operators ordered to satisfy $n_3\geq n_2\geq n_1$.
Again $\tilde{Q}_5$ is chosen to annihilate primary states, while $\hat{Q}_5$ is a function of $L_0$. It is easy to see from $\ell$-scaling of saddle point values of $\Delta, n$ that only $L_0^3$ term from $\hat{Q}_5$ contributes to the extensive part of free energy. Similarly to the case of $Q_3$, we drop $\tilde{Q}_5$ at leading  order in $c$. The rest is straightforward and can be generalized to all $Q_{2k-1}$. After integrating over $n$ the partition function can be reduced to the integral over $E=\Delta+n$,
\bea
\label{Zleadingc}
Z(\beta,\mu_{2k-1})=\int dE\, e^{{\mathcal L}_E},\quad 
{\mathcal L}_E=\pi\sqrt{\frac{2}{3} c E}-{\beta\over \ell}E-{\mu_3\over \ell^3}E^2-{\mu_5\over \ell^5}E^3-\dots.
\eea
It is easy to see than that free energy will depend on the inverse temperature $\beta$ and fugacities $\mu_i$ only through the combinations
\bea
\label{tdef}
F={c\,\pi^2\ell\over 6\, \beta}f_0(t_{2k-1}),\qquad t_{2k-1}=\left( {c\, \pi^2 \over 6\, \beta^2}\right)^{k-1} {\mu_{2k-1}\over \beta}.
\eea
Free energy $F$ admits perturbative expansion in $t_{2i-1}$ at any value of $c$, while expansion in $\mu_{2k-1}$ breaks down for large central charge. In principle $f_0$ is given by the algebraic equation specifying the saddle point $E^*$ of \eqref{Zleadingc},
\bea
\label{effectiveLE} {\mathcal L}_E&=&{c\,\pi^2\ell\over 6\, \beta}s_0,\qquad \qquad 
E^*={c\,\pi^2\ell\over 6\, \beta}e^*,\\ \label{effS}
s_0&=&2\sqrt{e}-e-t_3 e^2 -t_5 e^3-\dots,\\ 
f_0&=&s_0(e^*),\qquad\qquad\qquad  
\left.{\partial s_0\over \partial e}\right|_{e^*}=0.
\label{effF}
\eea
In practice perturbative expansion of $f_0$ is easier to recover iteratively. We notice that at leading order in $c$ the expectation values of $Q_{2k-1}$ are given by $\ell (E^*/\ell^2)^k$. In terms of the partition function \eqref{Zleadingc} this can be rewritten as a differential equality
\label{sec:infinitec}
\bea
\label{relation}
-\ell^{-1}\partial_{\mu_{2k-1}}\log Z=(-\ell^{-1}\partial_{\beta}\log Z)^k,
\eea
or in terms of variables $t_{2k-1}$,
\bea
\label{eq1}
(f_0+3t_3 \partial_{t_3}f_0+5t_5 \partial_{t_5}f_0+\dots)^2+ \partial_{t_3}f_0=0\ ,\\
\label{eq2}
(f_0+3t_3 \partial_{t_3}f_0+5t_5 \partial_{t_5}f_0+\dots)^3+ \partial_{t_5}f_0=0\ ,\\
\dots \nonumber
\eea
For the Taylor expansion of $f_0$ equations (\ref{eq1}),(\ref{eq2}), etc.~yield iterative relations which can be easily solved, 
\bea
f_0=1 - t_3 + 4 t_3^2 - 24 t_3^3 - t_5 + 12 t_3 t_5 - 132 t_3^2 t_5 + 9 t_5^2 - 
 234 t_3 t_5^2 - 135 t_5^3+\dots
\eea
An elegant expression for $f_0$ in terms of the perturbative series in $t_{2k-1}$ was given recently in \cite{maloney2}.

\subsection{GGE at Finite Central Charge}
\label{sec:finitec}
When the central charge is finite, extensive part of free energy acquires $1/c$ corrections, 
\bea
\label{fexpansion}
F={c\,\pi^2\ell\over 6\, \beta}\left(f_0+{f_1\over c}+{f_2\over c^2}+\dots\right).
\eea
Functions $f_k$ admit Taylor expansion in terms of  $t_{2k-1}$ \eqref{tdef}. 
For $k\geq 2$, $f_k$  depends only on $t_{2i-1}$, $i\geq k$, and  start with a term linear in $t_{2k-1}$. Thus, if expressed in terms of $\mu_{2k-1}$, extensive part of free energy is polynomial in $c$. 
To calculate $f_1, f_2, \dots$ we need to take into account additional contribution to the effective action, 
\bea
\label{fulleffectiveL}
e^{\tilde{\mathcal L}(c,\mu_{2k-1}/\ell^{2k-1},\Delta, n)}&\equiv&{1\over P(n)}\sum_{\{m\}=n} \langle m_i,\Delta|e^{-{\mu_3}\tilde{Q}_3-{\mu_5}\tilde{Q}_5-\dots}|m_i,\Delta\rangle. \label{summfull}
\eea

\subsubsection{$1/c$ corrections from $\tilde{Q}_3$}
\label{tildeQ3}
We start with the case when only $\mu_3\neq 0$, \eqref{summ}. 
It is easy to see that the operator $\tilde{Q}_3$ written in the basis \eqref{basis} is not more than linear in $c$ and $\Delta$, 
\bea
\label{tildeQ3decomposition}
\tilde{Q}_3=c\, \tilde{Q}_3^c +\Delta\, \tilde{Q}_3^\Delta+ \tilde{Q}_3^{(0)}.
\eea  
Matrices $\tilde{Q}_3^c, \tilde{Q}_3^\Delta$ have lower-triangular form with the diagonal elements being 
\bea
\ell^3\tilde{Q}_3^c |m_i,\Delta\rangle &=& \lambda |m_i,\Delta\rangle+\dots, \qquad  \quad \lambda=\frac{1}{6} \left(\sum_i m_i^3 -m_i \right),\\
\ell^3\tilde{Q}_3^\Delta  |m_i,\Delta\rangle &=& \nu |m_i,\Delta\rangle+\dots, \qquad \qquad \qquad  \nu=4\, n.
\eea
Because of the triangular form, $c\,\lambda+\Delta\, \nu$ are the eigenvalues of $c\,\tilde{Q}_3^c+\Delta\, \tilde{Q}_3^\Delta$. 
To estimate contribution of each term in \eqref{tildeQ3decomposition} toward free energy in the Appendix \ref{appendix:sum} we calculate  
\bea
\label{3sum}
\ell^3\langle \tilde{Q}_3\rangle_{\Delta,n}\equiv {\ell^3\over P(n)} \sum_{\{m\}=n} \langle m_i,\Delta| \tilde{Q}_3|m_i,\Delta\rangle =(a_0 c + b_0) n^2+4 \Delta n+O(\ell^3),\label{1pt}
\eea
in the limit of infinite $n$, assuming $n,\Delta \sim \ell^2$. The $n$-independent coefficients
\bea
\label{a0}
a_0={2\over 5},\qquad b_0=4.
\eea
We note that leading $n$-scaling  in \eqref{1pt}  is fixed to be $n^2$ lest the contribution of $\tilde{Q}_3$ diverge in the thermodynamic  limit $\ell\rightarrow \infty$.
At leading order in $c$ this is easy to see directly from the sum over Young tableaux in \eqref{3sum},
\bea
{1\over P(n)} \sum_{\{m\}=n} \left({1\over 6}\sum_i m_i^3\right) =a_0  n^2+O(n).
\eea
For large $n$ there are exponentially many  ways to represent $n$ as a sum of integers  with the typical partition consisting of $\sim\sqrt{n}$ terms with each term being of order $m_i\sim \sqrt{n}$. From here it immediately follows that $\sum_i m_i^3\sim \sqrt{n} \times n^{3/2}\sim n^2$.
After substituting  scaling of saddle-point values $\Delta^* \sim c\ell^2$, $n^*\sim \ell^2$ into \eqref{1pt} we find that both $c\, \tilde{Q}_3^c$ and $\Delta\, \tilde{Q}_3^\Delta$ contribute toward $f_1$
and potentially to $f_2$, while $\tilde{Q}_3^{(0)}$ contributes to $f_2$ only. So far we are interested only in $f_1$ we can simplify \eqref{summ} to be 
\bea 
\label{effectiveL}
&&e^{\tilde{\mathcal L}(c,\, \mu_3/\ell^3,\Delta, n)}\equiv {e^{-{\mu_3\over \ell^3} 4 n \Delta  }\over P(n)}\sum_{\{m\}=n} e^{-{\mu_3\over \ell^3} {c\over 6}\sum_i m_i^3  },\\
\label{effectiveaction3}
&&\tilde{\mathcal L}=-\mu_3/\ell^3\, 4\, n\, \Delta  -{a_0}\,  (c\, \mu_3/\ell^3) n^2-a_1\, (c\, \mu_3/\ell^3)^2  n^{7/2}-a_2\,(c\, \mu_3/\ell^3)^3 n^{5}+\dots\quad \label{Leffexpansion}
\eea
The expansion \eqref{Leffexpansion} assumes $\ell^3/(c\,\mu_3)\gg n\gg 1$. 
Sum over Young tableaux \eqref{effectiveL} provides a non-perturbative definition of $\tilde{\mathcal L}$. It is a simplified version of the sums appearing in  \cite{losev2005small}. A few first coefficients $a_i$ are calculated in the Appendix \ref{appendix:a}.

In principle effective action \eqref{Leffexpansion} together with \eqref{effL} completely determines $f_1(t_3)$ via maximization over $\Delta$ and $n$, but there is a direct way to obtain $f_1$ bypassing this step. We first notice that at leading order in $c$ effective action \eqref{effL} as a function of $\Delta$ reduces to \eqref{Zleadingc} and therefore saddle-point value of $\Delta^*$ is fixed by  \eqref{effS} independently of the value of $n^*$,
\bea
\Delta^*={c\pi^2 \ell^2\over 6\, \beta^2}\left(e^*+O(1/c)\right).
\eea
In case when only $t_3$ is ``turned on'' $\z$ can be found explicitly,
\bea
\nonumber
\z={(t_3-(t_3^3+27 t_3^4+3^{3/2}t_3^{7/2}\sqrt{2+27 t_3})^{1/3})^2\over 6 t_3^2 (t_3^3+27 t_3^4+3^{3/2}t_3^{7/2}\sqrt{2+27 t_3})^{1/3}}
=1-4t_3+28 t_3^2-240 t_3^3+2288t_3^4-23296 t_3^5+O\left(t_3^6\right)
\eea
Then leading $1/c$ correction to free energy can be found now by plugging $\Delta^*$ back into \eqref{effL} and keeping $O(c^0)$ terms,
\bea
e^{{\pi^2 \ell\over 6\beta}f_1}=e^{-{\pi\over 2}\sqrt{\Delta^*\over 6 c}}\sum_{\{m_i\}} e^{-{\beta\over \ell}n-6{\mu_3\over \ell^3}n \Delta^* -{\mu_3\over \ell^3}{c\over 6}\sum_i m_i^3}. \label{sumY}
\eea
Here $n=\sum_i m_i$ and the sum goes over all Young tableaux.  This sum can be calculated by rewriting \eqref{sumY} using free boson representation.\footnote{We thank Nikita Nekrasov for help with the following calculation.} Namely, each set $m_i$ will be parametrized by the set of integer numbers $r_k$, where $r_k$ is the number of times natural number $k$ appears in the set $m_i$. Then the sum over all sets $\{m_i\}$ is equivalent to the sum over all $r_k$ and 
\bea
\label{bosonic}
\sum_i m_i=\sum_{k=1}^\infty k\, r_k,\qquad \sum_i m_i^3=\sum_{k=1}^\infty k^3 r_k. 
\eea
For any coefficients $x,y>0$ we find 
\bea
\sum_{\{m_i\}} e^{-x \sum_i m_i -y \sum_i m_i^3}= \sum_{r_1=0}^\infty \sum_{r_2=0}^\infty \dots e^{-\sum_k r_k (x k+y k^3)}=\prod_{k=1}^\infty \left(1-e^{-x k -y k^3}\right)^{-1}.\ \ \  \label{product}
\eea
This  infinite product  can be consequently rewritten as an exponent of the sum of logarithms. Going back to \eqref{sumY} this gives 
\bea
\label{infsum}
{\pi^2 \ell \over 6\beta}f_1=-{\pi^2 \ell \over 6\beta}\sqrt{\z}-\sum_{k=1}^\infty \log\left(
1-e^{-{\beta\over \ell}(1+6 t_3 \z)k-{t_3\over \pi^2}\left({\beta\over \ell}\right)^3k^3}
\right).
\eea
The effective variables in \eqref{infsum} is the combination $\beta k/\ell$ and in the limit $\ell\rightarrow \infty$ the sum over $k$ can be substituted by an integral, yielding 
\bea
f_1&=&-\sqrt{\z}-{6\over \pi^2}\int_0^\infty dk \log\left(
1-e^{-(1+6t_3 \z )k -t_3{k^3/\pi^2}}
\right).
\eea
A few first terms in Taylor series expansion of $f_1$ can be easily calculated by expanding this expression in $t_3$,
\bea
f_1=-\frac{22 }{5}t_3+\frac{2096 }{35}t_3^2-\frac{4464}{5} t_3^3+\frac{82304 }{5}t_3^4+O\left(t_3^5\right).
\eea
This agrees with the perturbative calculation of \cite{maloney2}.
%
%

\subsubsection{$1/c$ corrections from $\tilde{Q}_5$}
Adding $\mu_5$ to consideration is straightforward. We notice that $\tilde{Q}_5$ is not more than quadratic in $c$ and $\Delta$,
\bea
\tilde{Q}_5=c^2 \tilde{Q}^{cc}_5 +c\, \Delta \tilde{Q}_5^{c\Delta}+\Delta^2 \tilde{Q}_5^{\Delta\Delta}+c\,\tilde{Q}_5^c +\Delta \tilde{Q}_5^\Delta+\tilde{Q}_5^{(0)},
\eea
and all three matrices $\tilde{Q}^{cc}$, $\tilde{Q}^{c\Delta}$, $\tilde{Q}_5^{\Delta\Delta}$ are lower-triangular in the  basis \eqref{basis}.  Their diagonal matrix elements are easy to calculate,
\bea
\ell^5\tilde{Q}_5^{cc} |m_i,\Delta\rangle &=& \alpha|m_i,\Delta\rangle+\dots, \quad  \alpha =\frac{1}{12} \left( \sum_i \frac{m_i^5}{6}  - \frac{5  m_i^3}{12} -\frac{m_i}{4} \right),\\
\ell^5\tilde{Q}_5^{c\Delta}   |m_i,\Delta\rangle &=& \delta |m_i,\Delta\rangle+\dots, \qquad \qquad \qquad  \delta=  \sum_i \frac{5}{6} m_i^3 -m_i ,\\
\ell^5\tilde{Q}_5^{\Delta \Delta}   |m_i,\Delta\rangle &=& \gamma |m_i,\Delta\rangle+\dots, \qquad \qquad \qquad  \gamma=12\, n.
\eea
Only these three matrices contribute toward $f_1$ via 
\bea
\label{effL5}
e^{\tilde{\mathcal L}(c,\, \mu_3/\ell^3,\, \mu_5/\ell^5, \Delta, n)}&\equiv& {e^{-{\mu_3\over \ell^3} 4 n\Delta-{\mu_5\over \ell^5}12 n \Delta^2}\over P(n)}\sum_{\{m\}=n} e^{-\left({c\mu_3\over 6\,\ell^3}+{5c\Delta \mu_5\over 6\,\ell^5}\right) \sum_i m_i^3  -{c^2\mu_5\over 72 \ell^5} \sum_i m_i^5}.
\eea
Next steps are completely analogous to the case with only $Q_3$. At leading order in $c$ saddle point value of $\Delta$ is fixed by 
\bea
\label{leading}
s_0&=&2\sqrt{e}-e-t_3 e^2-t_5 e^3,\\
f_0&=&s_0(\z),\qquad \left.{ds_0\over de}\right|_{\z}=0,\qquad \Delta^*={c\pi^2 \ell^2\over 6\, \beta^2}\z, \label{saddlet5}
\eea
while $f_1$ is given by 
\bea
\label{otvet}
f_1=-\sqrt{\z}-{6\over \pi^2}\int_0^\infty dk \log\left(
1-e^{-(1+6t_3 \z+15 t_5 (\z)^2)k -(t_3+5 t_5 \z){k^3/\pi^2}-{1\over 2}t_5 {k^5/\pi^4}}
\right).
\eea
A few first terms in Taylor series expansion are
\bea
f_1=-\frac{22 }{5}t_3- {302\over 21}t_5 +\frac{2096 }{35}t_3^2+\frac{14328}{35}t_3 t_5+\frac{51168}{77}t_5^2+\dots
\eea

Generalization to include higher charges is conceptually straightforward. The leading term $f_0$ is given by \eqref{effS} with   $s_0=2\sqrt{e}-e-\sum_{j=2}^\infty t_{2j-1} e^j$. We expect that all higher charges $\tilde{Q}_{2j-1}$ also have lower-triangular form at leading in order $c$, and therefore \eqref{fulleffectiveL} will reduce to an appropriate sum over Young tableaux, yielding an appropriate generalization of \eqref{otvet}.

\section{Comparison of GGE with Eigenstate Ensemble}
\subsection{Eigenstate Thermalization Hypothesis}
\label{sec:ETH}
Eigenstate Thermalization Hypothesis (ETH), developed in the works of Deutsch and Srednicki in the 90s \cite{deutsch1991quantum, srednicki1994chaos}  (also see \cite{shnirelman} for an earlier work on the subject) is the idea that individual energy eigenstate of a sufficiently complex ``chaotic'' many-body quantum system exhibits thermal properties. ETH provides a mechanism explaining   thermalization of isolated quantum  systems. Extensive numerical studies during last decade have supported the expectation that ETH is a general property of quantum many-body systems, unless the system exhibits an extensive number of conserved quantities \cite{rigol2008thermalization}. 

At the colloquial level the ETH promotes a highly excited energy eigenstate of a quantum system to the ``eigenstate ensemble'' stating that the latter can describe thermal properties of a quantum system in the same way as the conventional canonical and micro-canonical ensembles. The discrepancy between e.g.~expectation values of local quantities in these different ensembles then would be suppressed in the thermodynamic limit. Holographic CFTs are expected to be complex enough to exhibit thermalization starting from a  sufficiently excited pure state. On the dual gravity side this is a process of  black hole formation via gravitational collapse. Consequently holographic CFTs are expected to exhibit eigenstate thermalization, at least in some form.

A basic argument of  \cite{lashkari2018eigenstate} shows that CFTs can not satisfy  the conventional form of ETH when all sufficiently excited energy eigenstates are thermal. This is because in the latter case  all eigenstates would have to exhibit the same thermal properties: expectation value of a local operator $\mathcal O$ would be a function of effective temperature, i.e.~would depend on energy only
\bea
\label{eth}
\langle E|\mathcal O|E\rangle={\mathcal O}_{\rm eth}(\beta(E)).
\eea
Operator-state correspondence allows reformulation \eqref{eth} in terms of OPE coefficients. Then it can be easily shown that the primary states  and the descendants can not both satisfy \eqref{eth} with the same function ${\mathcal O}_{\rm eth}$.

It was proposed in \cite{dymarsky2018subsystem} that the non-trivial content of ETH is not the equivalence between individual eigenstates and thermal ensemble, but the equivalence of  individual eigenstates from a certain class with each other. Then the equivalence with the thermal ensemble would follow automatically, provided the original class of eigenstates is wide enough. This idea lead \cite{lashkari2018eigenstate} to propose the following  formulation of local ETH in CFTs: \eqref{eth} should apply to any local operator $\mathcal O$ but be limited only to primary (Virasoro-primary in 2d) states $|E\rangle$. In this formulation ${\mathcal O}_{\rm eth}(\beta)$ is not necessarily related to thermal expectation value of $\mathcal O$. For conformal theories in $d\geq 3$ it was further argued in \cite{lashkari2018universality} that the primary states dominate the microcanonical ensemble and therefore in the thermodynamic limit ${\mathcal O}_{\rm eth}(\beta)$ coincides with the conventional thermal expectation value of $\mathcal O$. The $d=2$ case is a subject of discussion below. 

In 2d we need to distinguish two cases: when  $\mathcal O$ is a Virasoro descendant of  identity or not. In the former case the local ETH \eqref{eth} is automatic -- the corresponding heavy-heavy-light OPE coefficient is fixed by Virasoro algebra and is a smooth function of $E$, dimension of the Virasoro primary state $|E\rangle$. In the latter case, when $\mathcal O$ is not a descendant of identity, there are no known explicit examples when \eqref{eth} is satisfied with a smooth ${\mathcal O}_{\rm eth}(\beta)$. There is a general expectation though that this is the case for certain large central charge theories, including holographic CFTs. In $d=2$ thermal expectation value of any such $\mathcal O$ is zero (because thermal cylinder is conformally flat). It  is therefore often assumed that in sufficiently complex 2d theories corresponding heavy-heavy-light OPE  is suppressed by the dimension of the heavy operator. Dominance of the vacuum conformal family or the ``identity block" in the OPE of two heavy primaries is an underlying assumption in many works on large central charge theories in the context of ETH and thermalization \cite{anous2016black,chen2016short,lin2016thermality,He2017txy,basu2017thermality,He2017vyf,lashkari2018eigenstate,faulkner2018probing,Guo2018pvi,hikida2018eth,romero2018cardy,brehm2018probing}. 

Leaving aside the behavior of  $\mathcal O$ outside of vacuum conformal family, below we focus on the case when $\mathcal O$ is a Virasoro descendant of identity.  Local ETH \eqref{eth} is automatic in this case, but function ${\mathcal O}_{\rm eth}(\beta)$ a priory has no interpretation in terms of thermal physics. 	A natural question then would be to compare ${\mathcal O}_{\rm eth}(\beta)$ with thermal expectation value of the operator $\mathcal O$,
\bea
\label{thermal}
{\mathcal O}_{\rm th}(\beta)={1\over Z(\beta)}{\rm Tr}( {\mathcal O}\, e^{-\beta H}). 
\eea																																
It was expected that at infinite central charge locally eigenstate is equivalent to the thermal ensemble \cite{fitzpatrick2014universality}, which is indeed the case, ${\mathcal O}_{\rm eth}={\mathcal O}_{\rm th}$.  At the same time ${\mathcal O}_{\rm eth}$ and ${\mathcal O}_{\rm th}$ do not match at the subleading order in $1/c$ \cite{lin2016thermality,He2017vyf,basu2017thermality,He2017txy,lashkari2018universality}.
A naive explanation of this discrepancy is that the ``eigenstate ensemble'' $|E\rangle$ has positive value of qKdV charges and hence should be compared not with \eqref{thermal} but with the full Generalized Gibbs Ensemble  
\bea
\label{gge}
{\mathcal O}_{\rm GGE}(\beta,\mu_i)={1\over Z(\beta,\mu_i)}{\rm Tr}( {\mathcal O}\, e^{-\beta H-\sum_k \mu_{2k-1} Q_{2k-1}}), 
\eea
with the fugacities $\mu_i$  chosen to match quantum numbers of $|E\rangle$. This is the comparison performed below. 																														
																									
\subsection{Comparison at infinite central charge}
From now on we restrict to the case when $|E\rangle$ is a heavy {\it scalar} Virasoro primary.  To achieve finite energy density and thus finite  effective temperature in the thermodynamic limit the dimension $E$ should scale as $\ell^2$, 
\bea
\ell\rightarrow  \infty, \qquad E/\ell^2={\rm fixed}. \label{thermodynamic}
\eea
A Virasoro descendant of identity $\mathcal O$ can be either a quasi-primary or  a total derivative. In the latter case by Lorentz invariance ${\mathcal O}_{\rm eth}={\mathcal O}_{\rm GGE}=0$. When $\mathcal O$ is a quasiprimary of dimension $2k$,  ${\mathcal O}_{\rm eth}$ and  ${\mathcal O}_{\rm GGE}$ are non-trivial in the thermodynamic limit \eqref{thermodynamic} only if $\mathcal O$ includes $\underbrace{(T\dots (TT))}_{\rm k\, times}$. One can always choose a basis at the level $2k$ such that a unique quasi-primary with a non-vanishing expectations values (\ref{eth},\ref{gge}) in the thermodynamic limit  is the density of KdV charge ${\mathcal O}=q_{2k-1}$ \cite{lashkari2018universality}, 
\bea
Q_{2k-1}\equiv \int_0^{\ell} du\, q_{2k-1}.
\eea
Thus to compare the eigenstate and generalized Gibbs ensembles  it suffices to compare 
\bea
\label{QE}
\langle E|q_{2k-1}|E\rangle={1\over \ell}\langle E|Q_{2k-1}|E\rangle =\left({ E\over \ell^2}\right)^k, \qquad  E,\ell \rightarrow +\infty,
\eea
and 
\bea
\label{QGGE}
\langle q_{2k-1}\rangle_{GGE}={1\over \ell}\langle Q_{2k-1}\rangle_{GGE}\equiv -{1\over \ell}\partial_{\mu_{2k-1}}\log Z,
\eea
for $k \geq 2$. For $k=1$ the equality between $\langle E|T|E\rangle$ and $\langle T\rangle_{GGE}$,
\bea
\label{effbeta}
E=\langle H\rangle_{GGE}
\eea
is the relation which defines effective temperature $\beta$ in terms of $E/\ell^2$.
The expectation values \eqref{QE} satisfy $\langle E|q_{2k-1}|E\rangle = \langle E|q_1|E\rangle^{k}$, $q_1\equiv T$. Therefore for the equality between the eigenstate and generalized Gibbs ensemble to hold it is necessary and sufficient for the GGE partition function to satisfy \eqref{relation}. This is the case at leading order in central charge for any values of  $\mu_{2k-1}$ as discussed in section \ref{sec:infinitec}. Hence we establish that for any descendant of identity $\mathcal O$, at leading order in central charge expectation value in a primary state is the same as in the generalized Gibbs ensemble for any choice of $\mu_{2k-1}$. This is consistent with the holographic interpretation
that at infinite central charge  classical black hole in $AdS_3$ is dual to \eqref{GGE} for any values of  $\mu_{2k-1}$ \cite{de2016remarks}.

\subsection{Discrepancy at finite central charge}
At finite central charge the relations \eqref{relation} are not automatically satisfied. Using the expansion \eqref{fexpansion}, at leading order in $1/c$ one finds from (\ref{eq1},\ref{eq2}),
\bea
\label{Eq1}
2(f_0+3t_3 \partial_{t_3}f_0+5t_5 \partial_{t_5}f_0+\dots)(f_1+3t_3 \partial_{t_3}f_1+5t_5 \partial_{t_5}f_1+\dots)+ \partial_{t_3}f_1=0,\\
\label{Eq2}
3(f_0+3t_3 \partial_{t_3}f_0+5t_5 \partial_{t_5}f_0+\dots)^2(f_1+3t_3 \partial_{t_3}f_1+5t_5 \partial_{t_5}f_1+\dots)+ \partial_{t_5}f_1=0,\\ 
\dots \nonumber
\eea
In principle one can hope these equations would specify a set of $t_{2k-1}$ such that 
(\ref{Eq1},\ref{Eq2},\dots) are satisfied, which would ensure equivalence between the eigenstate and generalized Gibbs ensemble at the first subleading order in $1/c$. It should be noted though that the equations (\ref{Eq1},\ref{Eq2},\dots) have no small parameter, because when written in terms of variables $t_{2k-1}$ they are $c$-independent. Thus one would need to know full function $f_1(t_3,t_5,\dots)$ to find a possible solution. In this sense the problem of matching fugacities $\mu_{2k-1}$ to a primary state is non-pertrubative, i.e.~it requires knowledge of free energy at all orders in $\mu_{2k-1}$, even for large $c$ \cite{lashkari2018universality}. 

Although we do not know the full function $f_1(t_3,t_5,\dots)$, a simple argument shows that a solution of \eqref{Eq1} and more generally of \eqref{relation} does not exist for finite but large $c$. 
Let us compare \eqref{QE} with \eqref{QGGE} for $Q_3$ by calculating the difference between the two, 
\bea
\label{discrepancy}
\ell^{-1}\left(\langle Q_3\rangle_{GGE}-\langle E|Q_3|E\rangle\right)=
\ell^{-1}\left(\langle \hat{Q}_3\rangle_{GGE}+\langle \tilde{Q}_3\rangle_{GGE}\right) -\left({E\over \ell^2}\right)^2.
\eea
Using explicit expression for $\hat{Q}_3$ \eqref{Q3} we find that in the thermodynamic limit the expectation value of $\hat{Q}_3$ is given by the saddle point,
\bea
\label{E2}
\ell^{-1}\langle \hat{Q}_3\rangle_{GGE} =\left({E^*\over \ell^2}\right)^2 +O(1/\ell). 
\eea
The saddle point value $E^*=\Delta^*+n^*$ is equal to the energy $E$ of state $|E\rangle$ due to \eqref{effbeta}. 
The values of $\Delta^*$, $n^*$  should be determined from the full effective action ${\mathcal L}+\tilde{\mathcal L}$ \eqref{effL}, \eqref{fulleffectiveL}. The full effective action is not known, but both $\Delta^*$ and $n^*$ scale with the system size as $\sim\ell^2$. Furthermore, up to $1/c$ corrections $\Delta^* \approx E^*=E$ and therefore $\Delta^*/\ell^2$ is a positive number in the thermodynamic limit. Using \eqref{discrepancy} and \eqref{E2} we find the discrepancy between the eigenstate and GGE expectation values to be 
\bea
\label{discrepancy2}
&&\ell^{-1}\left(\langle Q_3\rangle_{GGE}-\langle E|Q_3|E\rangle\right)=
\ell^{-1}\langle \tilde{Q}_3\rangle_{GGE}= \\
&&\qquad\qquad\qquad {e^{-\tilde{\mathcal L}(\Delta^*,n^*)}\over \ell\, P(n^*)}\sum_{\{m\}=n^*} \langle m_i,\Delta^*|\tilde{Q}_3 e^{-{\mu_3}\tilde{Q}_3-{\mu_5}\tilde{Q}_5-\dots}|m_i,\Delta^*\rangle. \nonumber
\eea
So far we are interested in the leading in $c$ behavior of  \eqref{discrepancy2} we can substitute $\tilde{Q}_3$ by a lower-triangular matrix $c\,\tilde{Q}_3^c+\Delta^* \tilde{Q}_3^\Delta$, $\tilde{Q}_5$ can be substituted by the lower-triangular $c^2\tilde{Q}_5^{cc}+c\,\Delta^* \tilde{Q}_5^{c\Delta}+ (\Delta^*)^2 \tilde{Q}_3^{\Delta\Delta}$, and so on. Then average of $\Delta^* \tilde{Q}_3^\Delta$ is simply equal to $4\Delta^* n^*/\ell^3$, while the average of  $c\,\tilde{Q}_3^c$ can be rewritten as an average over Young tableaux,
\bea
\nonumber
\langle \tilde{Q}_3^c\rangle_{GGE}={e^{-\tilde{\mathcal L}(\Delta^*,n^*)}\over 6\, \ell^3 P(n^*)}\sum_{\{m\}=n^*}\!\!\left(\sum_i m_i^3\right)\! e^{-{\mu_3\over \ell^3} \left({c\over 6}\sum_i m_i^3+4\Delta^* n^*\right)-{\mu_5\over \ell^5}\left({c^2\over 72}\sum_i m_i^5+\dots\right)-\dots}+O(c^0).
\eea
In the thermodynamic limit this is equal to $a\,(n^*)^2/\ell^3$, where $a$ is some non-negative function of $\mu_3,\mu_5,\dots$. When all $\mu_{2k-1}=0$, it reduces to $a_0=2/5$, \eqref{a0}. Thus we finally have 
\bea
\label{disc}
\ell^{-1}\left(\langle Q_3\rangle_{GGE}-\langle E|Q_3|E\rangle\right)={a\, (n^*)^2+4\Delta^* n^*\over \ell^4}+O(c^0).
\eea
The important point here is that the discrepancy \eqref{disc} can be zero if and only if $n^*=0$. The value of $n^*$ 
can be interpreted as the effective level of Virasoro descendants which give main contribution to the partition function. It is a priory expected that the discrepancy between the primary state and the GGE would vanish if $n^*=0$, i.e.~if primaries dominate the partition sum. The non-trivial result here is that  \eqref{disc} vanishes only if $n^*=0$. 

It is easy to see that primaries do not dominate the  partition sum because of the factor $P(n)$ accounting for the exponential growth of the number of descendants with $n$. Using \eqref{effL}, \eqref{effectiveL}, \eqref{Leffexpansion}, \eqref{effL5} the effective action for $n$ can be rewritten in terms of the variable $\n=n/\ell^2$ which remains finite in the thermodynamic limit, 
\bea
\label{eL}
{\mathcal L}_{\rm eff}(\n)&=&\ell\left(\pi\sqrt{2\n\over 3}-L\right),\\
L&=& (\beta+6\mu_3 \Delta^*/\ell^2+\dots)\n +   \mu_3(1+c a_0+b_0+\dots) \n^2 +O(\n^{5/2}). \nonumber
\eea
Since $L(\n)$ admits an expansion in powers of $\n$ starting from one, $\n=0$ can not be a solution of $\partial {\mathcal L}_{\rm eff}/\partial \n=0$ and the maximum of ${\mathcal L}_{\rm eff}$ is achieved at positive $\n=n^*/\ell^2$. From here it follows that for large $c$ both $\langle Q_3\rangle_{GGE}$ and $\langle E|Q_3|E\rangle$ scale as $c^2$, while their difference is non-zero and scales as $c$.
This proves \eqref{QE} and \eqref{QGGE} for $Q_3$ are always different for  large but finite $c$ for any values of $\mu_{2k-1}$.  A similar argument would also apply to $Q_5$ and higher charges.

\section{Discussion}
In this paper, we studied Generalized Gibbs Ensemble \eqref{GGE} of 2d CFTs in the limit when the size of the spatial circle $\ell$ goes to infinity. We have observed that  qKdV charges $Q_{2k-1}$ can be split into a sum of two terms:  first term $\hat{Q}_{2k-1}$ survives in the infinite $c$ limit and is a function of $L_0$, while  second term $\tilde{Q}_{2k-1}$ is $1/c$ suppressed and is positive semi-definite. Furthermore leading in $1/c$ expansion part of $\tilde{Q}_{2k-1}$ 
 assumes a simple lower-triangular form, if written in the conventional basis \eqref{basis}. 
Using this result we have explicitly calculated extensive contribution to free energy in the infinite central charge limit (\ref{tdef},\ref{effS},\ref{effF}), and reduced calculation of leading  $1/c$ correction to a sum over Young tabluex. The latter can be computed explicitly, provided the expression for the qKdV charge in terms of Virasoro algebra generators is known. We have calculated this correction for the first two qKdV charges $Q_3,Q_5$  \eqref{otvet}. Our results suggest adding higher qKdV charges at leading $1/c$ order will be straightforward, and we hope to report $f_1(t_3,t_5,t_7,\dots)$ which includes higher fugacities soon. Since the first two orders in $1/c$ expansion turned out to be easily tractable,  next natural question would be to understand the structure of the GGE partition function at the $1/c^2$ order. We leave this task for the future. Another important direction is to calculate sub-extensive corrections to free energy, even at leading order in $c$.  The logarithmic $\ln\ell$ and $O(\ell^0)$  terms would follow from the saddle point approximation of \eqref{effL}, while higher order terms in $1/\ell$ expansion will require more work.

Leading $1/c$ corrections to the GGE free energy calculated in this paper are of interest from the point of view of AdS/CFT correspondence. 
They encode quantum corrections on the gravity side and contain information about individual black hole microstates \cite{maloney2010quantum}.  
While from the point of view of local probes the heavy states  are thermal at infinite $c$ limit \cite{fitzpatrick2014universality,asplund2015holographic}, this is no longer the case at finite $c$. In this paper we compared a heavy Virasoro primary state with the GGE and found that the two are always locally distinguishable  no matter what the values of the qKdV fugacities are. 
 At the technical level this comes from the split of $Q_3$ into $\hat{Q}_3$ and $\tilde{Q}_3$. The expectation value of $\hat{Q}_3$ is the same in the primary eigenstate and the GGE for any values of the fugacities, but the expectation value of $\tilde{Q}_3$ is different. The operator $\tilde{Q}_3$ annihilates the primary state, while its expectation value in the GGE at leading order in $c$ is strictly positive. A similar result applies to higher charges as well.  
This mismatch between the eigenstate and the GGE is surprising. Large central charge theories are expected to be chaotic and thermalize \cite{asplund2015entanglement}, in full consistency with holographic expectations. Yet, we find there are initial configurations labeled by the macroscopic (extensive) values of qKdV charges which fail to thermalize i.e. be described by the GGE upon equilibration. 

The breakdown of ergodicity observed in this paper emphasizes the question of content and meaning of the Eigenstate Thermalization Hypothesis in 2d CFTs. First, it should be noted that ETH is always satisfied on average, i.e.~when \eqref{eth} is averaged over exponentially many eigenstates with similar energy. In fact the same applies to the off-diagonal matrix elements, on average they are exponentially small and described by a smooth form-factor. These is true for any theory, or in fact any system with a sufficiently large Hilbert space, and can be seen as a consequence of typicality. Correctness of ETH on average was confirmed for 2d CFT recently in \cite{kraus2017cardy,brehm2018probing,romero2018cardy,hikida2018eth}. As expected the obtained results are valid for any value of central charge, which confirms its kinematic Hamiltonian-independent origin. Arguably, the non-trivial content of ETH should be stronger and apply only to a subclass of chaotic theories. Second, the conventional understanding of Eigenstate Thermalization suggests that (certain) individual energy eigenstates should be compared directly with a thermal ensemble or its kin. We would like to argue for 2d CFTs this is not the correct approach.  
As was discussed in section \ref{sec:ETH}, primary energy eigenstates are  indeed thermal in the sense that \eqref{eth} is equal to \eqref{thermal}, but only at infinite $c$,  while this equality immediately breaks already at first subleaiding $1/c$ order  \cite{lin2016thermality,He2017vyf,basu2017thermality,He2017txy,lashkari2018universality}. A natural generalization of this idea would be to require the primaries to match the GGE. It was shown in this paper to  fail at finite $c$ as well. Besides the discrepancies at the level of local observables at finite $c$, there are other indications that the conventional form of ETH does not apply to primary states. As a non-local probe of thermal behavior one can consider Renyi entropy of an interval, when the size of the interval scales with $\ell$. In this limit eigenstate and thermal Renyi entropies are different even when ETH is obeyed, but in the latter case  the expression for the eigenstate entropy was conjectured in  \cite{dymarsky2018subsystem} in terms of the density of states. This conjecture was later studied and verified numerically in \cite{lu2017renyi} for non-integrable spin-chains. At the same time this conjectural form fails badly for  large $c$ conformal theories \cite{faulkner2018probing}. Another potentially related result is the breakdown of the KMS conditions by a heavy primary state \cite{fitzpatrick2016conformal}. 
All this strongly suggests that matching individual primary sates directly to a thermal or generalized thermal ensemble is incorrect. Maybe then the focus should be on the descendants?  A simple qualitative argument shows that heavy descendants also can not be thermal. If the state has energy $E=\Delta+n$,  $n\gg \Delta$,  through Virasoro algebra its properties are completely fixed in terms of $|\Delta\rangle$, which at least at infinite $c$, matches thermal state at wrong temperature. Non-thermal behavior of heavy descendant states was recently  rigorously  shown in \cite{Guo2018pvi}.

To conclude, we suggest that the nontrivial content of ETH in 2d CFTs is that for heavy primary states function ${\mathcal O}_{\rm eth}$ \eqref{eth} is a smooth function of energy, without it being directly related to thermal physics \cite{lashkari2018eigenstate}. In this sense our findings do not signal violation of ETH, 
in fact, we have explained in section \ref{sec:ETH} that for local probes from the vacuum conformal family this version of ETH is trivially satisfied. For this version of ETH to be physically interesting it should be established also for the probes outside of the vacuum family by demonstrating that at least in certain large $c$ theories all heavy-heavy-light OPE coefficients are smooth function of the heavy operator's dimension, or small. This is a challenging task without an immediate ways to approach it. Our point here  is that answering this question is crucial to understand if 2d CFTs exhibit an additional structure of heavy energy eigenstates beyond those imposed by symmetries and typicality.


\appendix
\section{Computation of $a_0$, $b_0$ }
\label{appendix:sum}
In this section we show how to calculate 
\bea
\ell^3\langle \tilde{Q}_3\rangle_{\Delta,n}\equiv {\ell^3\over P(n)} \sum_{\{m\}=n} \langle m_i,\Delta| \tilde{Q}_3|m_i,\Delta\rangle = (a_0 c + b_0) n^2+4 \Delta n+O(\ell^3),
\eea
in the limit when $\Delta,n\sim \ell^2$. An analogous calculation can be also found in \cite{maloney1}.
Recall that $\tilde{Q}_3 = 2\sum L_{-k} L_k$. The expectation value $\langle \tilde{Q}_3\rangle_{\Delta,n}$ can be thought of as trace $q^{- n - \Delta } \Tr (q^{L_0} \tilde{Q}_3)$ over a subspace with fixed $n$ and $\Delta$ spanned by the states \eqref{basis}. Using cyclic property of trace  one can easily get \cite{Apolo2015Q3}

\bea
\label{shenanigans}
\begin{split}
\langle L_{-k} L_k \rangle_{\Delta,n}= q^k \langle L_{k} L_{-k} \rangle_{\Delta,n} = \frac{q^k}{1 - q^k} \langle [L_{k},  L_{-k} ] \rangle_{\Delta,n} =\\ = \frac{q^k}{1 - q^k} \left( 2k \langle L_0 \rangle_{\Delta,n} + \frac{c}{12}(k^3 - k)\right),
\end{split}
\eea
where $q = e^{-\frac{\beta}{\ell}}$. Summing \eqref{shenanigans} over $k$  one obtains

\bea
\langle \tilde{Q}_3\rangle_{\Delta,n}\ = 4 \sigma_1 \langle L_0 \rangle_{\Delta,n} + \frac{c}{6} (\sigma_3 - \sigma_1),
\eea
where 
\bea 
\sigma_1 = \sum_k \frac{k q^k}{1 - q^k}, \\ \sigma_3 = \sum_k \frac{k^3 q^k}{1 - q^k}.
\eea
Expectation value $ \langle L_0 \rangle_{\Delta,n}$ is just equal to $n + \Delta$ and the sums $\sigma_1$, $\sigma_3$ can be evaluated in thermodynamic limit by replacing the sum over $k$ by an integral $\sum_k  \rightarrow \int dk$. In this limit
\bea
\sigma_1 \rightarrow \frac{\ell^2 \pi^2}{6 \beta^2}, \\ \sigma_3 \rightarrow \frac{\ell^4 \pi^4}{15 \beta^4}.
\eea
The final step is to recall that if we sum $\Tr (q^{L_0} \tilde{Q}_3)$ over $n$ main contribution will be given by a particular saddle point  value $n = \frac{\ell^2 \pi^2}{6 \beta^2}$.  Thus substituting $\frac{\ell^2 \pi^2}{6 \beta^2}$ by $n$ we get
\bea
\langle \tilde{Q}_3\rangle_{\Delta,n}\ = 4 n (n + \Delta) + \frac{2}{5} c\, n^2.
\eea
As a result, $a_0 = \frac{2}{5}$ and $b_0 = 4$. It is straightforward but tedious to generalize the above calculation to higher orders to obtain $a_1$, $a_2$ etc.~from \eqref{effectiveaction3}.

\section{Effective action at $1/c$}
\label{appendix:a}
In this section we show how to calculate the effective action  (\ref{effectiveaction3}), which is a partition function decorated by the first non-trivial qKdV charge $Q_3$ restricted to a particular large descendant level $n$,
\bea 
&&e^{\tilde{\mathcal L}(c,\, \mu_3/\ell^3,\Delta, n)}\equiv {e^{-{\mu_3\over \ell^3} 4 n \Delta  }\over P(n)}\sum_{\{m\}=n} e^{-{\mu_3\over \ell^3} {c\over 6}\sum_i m_i^3  }.
\eea
We remind the reader that first $1/c$ correction to free energy $f_1$ calculated in section \ref{tildeQ3} is given by (compare with  \eqref{sumY})
\bea
e^{{\pi^2 \ell\over 6\beta}f_1}=e^{-{\pi\over 2}\sqrt{\Delta^*\over 6 c}}\sum_{n} P(n)\, e^{-{\beta\over \ell}n-{\mu_3\over \ell^3}2n \Delta^* +\tilde{\mathcal L}(c,\, \mu_3/\ell^3,\Delta^*, n)}.
\eea
To calculate $\tilde{\mathcal L}$ we introduce an auxiliary partition function of the same kind, which depends on arbitrary parameter  $x$, 
\bea
e^{{\rm F}(x)}=\sum_{n} P(n)\, e^{-{x\over \ell} n+{\mu_3\over \ell^3}4 n \Delta +\tilde{\mathcal L}(c,\, \mu_3/\ell^3,\Delta, n)}.
\eea
In the limit of large $\ell$ this sum is saturated at some saddle point $n^*$,
\bea
\label{freeenergyax}
{\rm F}(x)=\pi \sqrt{2 n^*\over 3} -{x\over \ell}n^* +{\mu_3\over \ell^3}4 n^* \Delta+\tilde{\mathcal L}(c,\, \mu_3/\ell^3,\Delta, n^*),
\eea
which is related to $x$ and $\mu_3$ via $n^*=-\ell{dF\over dx}$.
``Free energy'' ${\rm F}$ can be calculated exactly the same way $f_1$ was calculated in section \ref{tildeQ3}, 
\bea
\label{Fex}
{\rm F}(x)=-\ell \int_0^\infty dk \ln\left(1-e^{-x k -{c\over 6}\mu_3 k^3}\right).
\eea
After changing the integration variable $k\rightarrow k/x$ and expanding in $\mu_3$ one finds
\bea
\label{Fnum}
{\rm F}(x)={\ell \over x}\left(\frac{\pi ^2}{6}-\frac{\pi ^4 c \mu_3}{90 x^3}+\frac{2 \pi ^6 c^2 \mu_3^2}{189 x^6} -\frac{4 \pi ^8 c^3 \mu_3^3}{135 x^9} +\frac{40 \pi ^{10} c^4 \mu_3^4}{243 x^{12}}+\dots\right)
\eea
Using the explicit expression \eqref{Fex} we can relate $n^*$ and  $x$,
\bea
{n^*\over \ell^2}={d\over dx} \int_0^\infty dk \ln\left(1-e^{-x k -{c\over 6}\mu_3 k^3}\right),
\eea
and solve this equation perturbatively with respect to $x$ by expanding in $\mu_3$,
\bea
x=\frac{\pi \ell}{\sqrt{6  n^*}}-\frac{4 c \mu_3 n^* }{5\ell^2}+\frac{144 \sqrt{6} c^2 \mu_3^2 (n^*)^{5/2}}{25 \pi \ell^5}
-\frac{20736 c^3 \mu_3^3 (n^*)^4}{25 \pi ^2 \ell^8}+
\frac{4762368 \sqrt{6} c^4 \mu_3^4 (n^*)^{11/2}}{125 \pi ^3 \ell^{11}}+\dots \nonumber
\eea
Plugging this back into \eqref{freeenergyax}  and finally renaming $n^*$ into $n$ gives (compare with \eqref{effectiveaction3})
\bea
&&{\mu_3\over \ell^3}4 n \Delta+\tilde{\mathcal L}(c,\, \mu_3/\ell^3,\Delta, n)=-\pi \sqrt{2 n\over 3}+{x(n)\over \ell}n+{\rm F}(x(n))=\\
&-&\frac{2}{5}\,  (c\, \mu_3/\ell^3) n^2+\frac{288\sqrt{6}}{175 \pi}\, (c\, \mu_3/\ell^3)^2  n^{7/2}  - \frac{20736}{125\pi^2}\,(c\, \mu_3/\ell^3)^3 n^{5} + \frac{732672\sqrt{6}}{125 \pi^3 }\,(c\, \mu_3/\ell^3)^4 n^{13/2}    \dots \nonumber
\eea
%
Generalization to include $Q_5$ and higher charges would be tedious but straightforward.

\acknowledgments
We would like to thank Alexander Avdoshkin, Shouvik Datta, Jared Kaplan, Alex Maloney, Gim Seng Ng, Vasily Pestun, Simon Ross, Ioannis Tsiares, Ilya Vilkoviskiy and especially Nikita Nekrasov for discussions. This work was supported by a grant of the Russian Science Foundation (Project No.~17-12-01587).



\bibliographystyle{JHEP}
\bibliography{GGE}

%
%
%
%
%
%
%
%
\end{document}